\documentclass[12pt,aps]{revtex4}
\usepackage{amsmath}
\usepackage{graphicx}

\begin{document}

\title{Potential problems with interpolating fields}
\author{Michael C. Birse}
\affiliation{Theoretical Physics Division, School of Physics and Astronomy,
The University of Manchester, Manchester, M13 9PL, UK}

\begin{abstract}
A potential can have features that do not reflect the dynamics of the
system it describes but rather arise from the choice of interpolating
fields used to define it. This is illustrated using a toy model of 
scattering with two coupled channels. A Bethe-Salpeter amplitude is
constructed which is a mixture of the waves in the two channels. 
The potential derived from this has a strong repulsive core, which 
arises from the admixture of the closed channel in the wave function
and not from the dynamics of the model. 
\end{abstract}

\maketitle

\section{Introduction}

In their work on two-baryon systems in lattice QCD, members of the HAL-QCD
collaboration have extracted baryon-baryon potentials from their simulations
\cite{iah07,niah09,ahi10,hal10,hal11,miah11,doi11}. The method relies on the
construction of a Bethe-Salpeter amplitude or ``wave function" -- 
a matrix element of two interpolating fields with the quantum numbers of 
the baryons involved. This amplitude is then inserted into a Schr\"odinger
equation in order to deduce the baryon-baryon potential that would generate it.
More recently, the same idea has been used to extract short-distance potentials 
through the operator product expansion of perturbative QCD 
\cite{abw10a,abw10b,abw11,abw12}. A review of the approach can be found in 
Ref.~\cite{aoki11}. 

One interesting feature of the resulting nucleon-nucleon potential is that it 
possesses a repulsive core \cite{iah07}, very reminiscent of those in many of
the potentials traditionally used in nuclear physics, such as Argonne $v_{18}$ 
\cite{wss95}. When extended to three flavours of quark with SU(3) symmetry, 
repulsive cores are seen in all channels except the SU(3) singlet \cite{hal10}. 
The repulsion is particularly strong in the symmetric octet channel, where no 
long-range attraction is seen. 

This pattern is qualitatively similar to that seen in constituent quark models 
(for reviews, see Refs.~\cite{oy84,osy00,fsn07}). For two octet baryons,
the symmetric octet channel is forbidden by the Pauli principle at small 
separations, leading to strong repulsion. Other channels are not forbidden and 
so, as well as antisymmetrisation, the form of the interaction between the 
quarks is needed, as discussed by Oka \cite{oka12}. In quark models based 
on magnetic gluon exchange, short-distance repulsion is found in all channels 
except the singlet.

The HAL-QCD approach has been criticised, for example in Refs.~\cite{dos07,bos08}, 
because a potential is not an observable in either experiments or lattice 
simulations. In the lattice case, the energies of two-baryon states are 
observables and these can be related, via L\"uscher's formula \cite{lu91}, 
to experimental observables, namely, phase shifts. In the HAL-QCD approach, this 
information on the physical phase shifts is encoded in the tails of the 
Bethe-Salpeter amplitudes outside the interaction region. In contrast, the short-distance forms of these amplitudes, and hence the deduced potentials are dependent 
on the choice of interpolationg field used to define them. 

More recently, the HAL-QCD collaboration have noted that the times at which they determine their potentials may not be sufficiently large
for a single state to dominate the correlator used to extract the 
Bethe-Salpeter amplitude. They have therefore introduced a version of their approach based on a time-dependent Schr\"odinger equation
\cite{hal11,hal12,ish12,kidah13}. A potential issue with this is 
that, since a number of energy eigenstates contribute to the 
correlator, there is no single phase shift determining the form
of Bethe-Salpeter amplitude at large separations. It is thus not obvious that the potentials determined from these amplitudes are
``anchored" to physical observables.

In this short note, I use a toy model of scattering with two coupled channels
to illustrate how a potential constructed in this way can develop features 
that reflect the choice of interpolating fields, and not the actual dynamics.
In particular an admixture of a closed channel can lead to a strong repulsive 
core in the derived potential that does not correspond to any aspect of the 
potentials in the original model. Finally, I comment on how this model may be able to check whether the time-dependent version of the 
HAL QCD approach leads to potentials that reproduce the correct
phase shifts.

\section{Coupled-channel model}

The toy model I use to illustrate issues with interpreting potentials from 
Bethe-Salpeter wave functions is two-body scattering in one dimension with 
two coupled channels. The lower-energy channel can be thought of as corresponding
to two nucleons in their ground states; the other to one ground-state nucleon and
one N$^*$ with the same quantum numbers. I shall refer to the channels as 
``NN" and ``NN$^*$".
The particles in the NN channel interact through an attractive square-well 
potential similar to the long-range attractive force between nucleons. 
For simplicity I take the potential in the NN$^*$ channel to be zero.
The N$^*$ has an excitation energy $\Delta$ and so the NN$^*$ channel is 
closed for energies below $\Delta$. The two channels are coupled by a contact 
interaction of strength $g$.

While this model is very much a caricature of real nucleon-nucleon scattering, 
I believe that it retains enough aspects of the real system to illustrate why 
one should be wary about taking seriously short-distance features.

The wave functions for the relative motion of the particles in the two channels, 
$\psi_0(x)$ and $\psi_1(x)$, satisfy the coupled equations,
\begin{eqnarray}
\left[-\,\frac{1}{M}\,\frac{d^2}{dx^2}+V(x)-E\right]\psi_0(x)
+2g\,\delta(x)\,\psi_1(x)&=&0,
\cr
\noalign{\vskip 5pt}
\left[-\,\frac{1}{M}\,\frac{d^2}{dx^2}+\Delta-E\right]\psi_1(x)
+2g\,\delta(x)\,\psi_0(x)&=&0,
\end{eqnarray}
where
\begin{equation}
V(x)=\left\{\begin{array}{ll}
-V_0&\quad\mbox{for}\quad |x|<a\cr
\noalign{\vskip 5pt}
0&\quad\mbox{for}\quad |x|>a.
\end{array}\right.
\label{eq:coupled}
\end{equation}

A general interpolating field for the ``nucleon" in this model is just
a linear combination of the N and N$^*$ fields. A Bethe-Salpeter amplitude 
$\Psi(x)$ can be defined for any coupled-channel solution by first building 
a second-quantised state representing the solution and then taking the matrix 
element of a pair of interpolating field operators between this state and the 
vacuum. The result is just a linear combination of the two wave functions, 
\begin{equation}
\Psi(x)=\psi_0(x)+p\,\psi_1(x).
\end{equation}
Since it is not required for the determination of the potential, I have not 
specified the overall normalisation of this quantity; only the coefficient
$p$ of the N$^*$ admixture is relevant.

The (even-parity) solutions of the coupled equations (\ref{eq:coupled}) 
have the forms
\begin{eqnarray}
\psi_0(x)&=&\left\{\begin{array}{ll}
A\cos(K|x|)+B\sin(K|x|)&\quad\mbox{for}\quad |x|<a\cr
\noalign{\vskip 5pt}
C\cos(k|x|)+D\sin(k|x|)&\quad\mbox{for}\quad |x|>a,
\end{array}\right.\cr
\noalign{\vskip 5pt}
\psi_1(x)&=&\exp(-\alpha|x|),
\end{eqnarray}
where
\begin{eqnarray}
k^2&=&ME,\cr
K^2&=&M(E+V_0),\cr
\alpha^2&=&M(\Delta-E).
\end{eqnarray}
Again the overall normalisation is irrelevant, and so I have arbitrarily
taken $\psi_1(0)=1$.

Matching the solutions at the edge of the square well and at the origin,
where the $\delta$-function acts, leads to the boundary conditions,
\begin{eqnarray}
\psi_0(a-)&=&\psi_0(a+),\cr
\psi_0^\prime(a-)&=&\psi_0^\prime(a+),\cr
\psi_0^\prime(0)&=&Mg\,\psi_1(0),\cr
\psi_1^\prime(0)&=&Mg\,\psi_0(0).
\end{eqnarray}
These give rise to a set of linear equations for the coefficients,
$A$, $B$, $C$, $D$:
\begin{eqnarray}
A\cos(Ka)+B\sin(Ka)&=&C\cos(ka)+D\sin(ka),\cr
-KA\sin(Ka)+KB\cos(Ka)&=&-kC\sin(ka)+kD\cos(ka),\cr
KB&=&Mg,\cr
-\alpha&=&MgA,
\end{eqnarray}
which can be solved straightforwardly. The resulting wave functions at energy
$E$ will be denoted $\psi_{0,1}(x;E)$. From these, we can construct 
the Bethe-Salpeter amplitude $\Psi(x;E)$.

Following the approach outlined in the introduction, I now use $\Psi(x;E)$
to define a potential by demanding that it satisfy the  Schr\"odinger equation,
\begin{equation}
\left[-\,\frac{1}{M}\,\frac{d^2}{dx^2}+V_{BS}(x;E)-E\right]\Psi(x;E)=0.
\end{equation}
The resulting potential is
\begin{equation}
V_{BS}(x;E)=\frac{1}{M\Psi(x;E)}\,\frac{d^2\Psi}{dx^2}+E.
\label{eq:bspotential}
\end{equation}

\section{Results}

I present results here for a representative parameter set, $MV_0=1/a^2$, 
$M\Delta=6/a^2$ and $M\,g=6/a$, and I take $a=1$ for convenience in plotting
the results. Fig.~1 shows the zero-energy potential $V_{BS}(x;0)$ 
for a range of values of the mixing parameter in the interpolating field. 
Note that I have not plotted the potential for $x>a$, where it is essentially 
zero, nor have I shown the $\delta$ function that is present at the origin.
\begin{figure}[ht]
\includegraphics[width=10cm,keepaspectratio,clip]{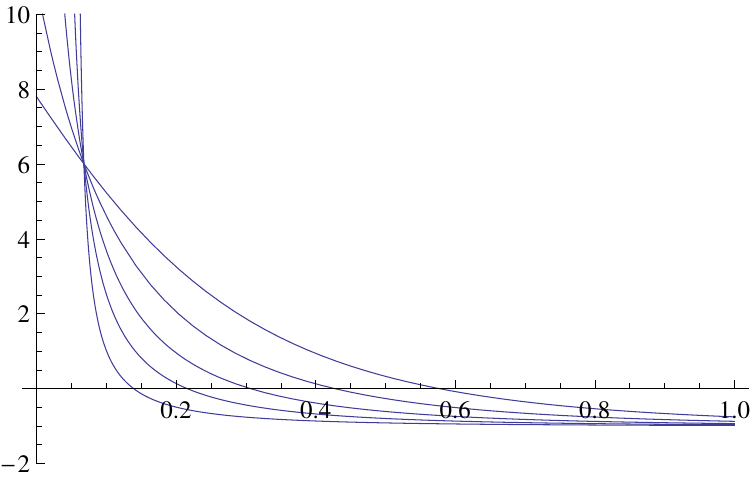}
\caption{The zero-energy potential $MV_{BS}(x;0)$ for 
$p=0.1$ (narrowest``core"), 0.25,  0.5, 1, 2 (widest).}
\end{figure}

A couple of features are worth noting about these potentials. First, the 
long-range attraction in the NN channel, $V(x)=-V_0$, can be seen as $x$ 
approaches 1. This similar to the way that the pion-exchange tail can be 
seen in the potentials extracted by the HAL QCD collaboration \cite{iah07}. 
The second feature is the apparent repulsive ``hard core" that is present
in all cases for $x\lesssim 0.5$. This comes entirely from the closed 
NN$^*$ channel, which makes a large positive contribution to $V_{BS}(x;0)$
for $x\lesssim 1/\alpha$ because $d^2\psi_1(x)/dx^2\simeq M\Delta\psi_1(x)$ 
there. It has nothing to do with the NN-channel wave function, as 
demonstrated by its strong dependence on the choice of interpolating 
field. In contrast, the long-range attraction (where it is not swamped by
the core) is independent of the choice of field.

\begin{figure}[ht]
\includegraphics[width=10cm,keepaspectratio,clip]{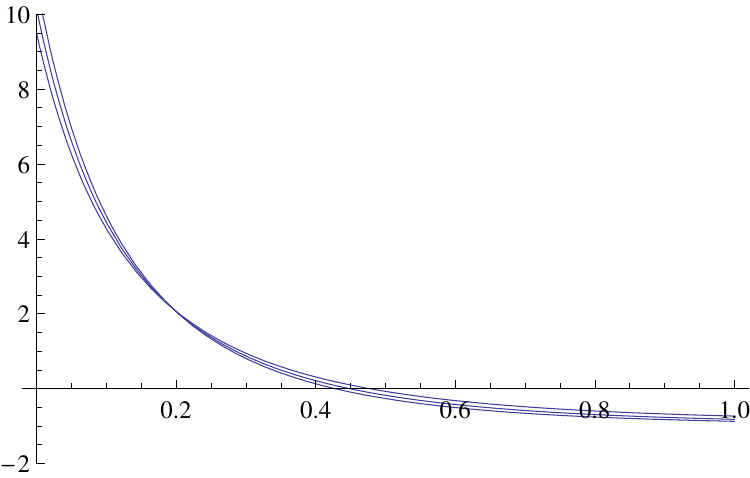}
\caption{The potential $MV_{BS}(x;E)$ for $ME=0$, $1$, $2$, all with $p=1$.}
\end{figure}
At low energies, the core in this potential is only weakly dependent on the 
energy $E$, as can be seen in Fig.~2. Like the HAL-QCD potentials, this has the appearance of something that could be well approximated by a local potential.
However this is just a reflection of the fact that the strength and range
of the core arise from the closed-channel wave function $\psi_1(x)$, and 
so are controlled by $g$ and $\Delta$ (and $p$ of course), so long as
$E$ lies well below the excitation energy $\Delta$.

\section{Conclusion}

The results in the previous section illustrate how a potential can have features 
that arise from the interpolating fields used to define it. In particular they 
show how coupling to closed channels can lead to repulsive cores in potentials 
derived from Bethe-Salpeter amplitudes. Similar repulsive cores seen in the 
HAL-QCD collaboration's analyses of their lattice simulations \cite{iah07,niah09,ahi10,hal10,hal11,miah11,doi11} could therefore be results of 
their choice of interpolating fields, rather than indications of similar 
dynamical mechanisms to those of the constituent quark model.

Since the same interpolating fields are used in all channels, one might argue
that the absence of a repulsive core in the SU(3)-singlet channel is evidence
against these cores being artefacts of the choice of field. However the strength 
of the core is dependent on the size of the coupling to the closed channel and,
for baryons in the SU(3) limit, this coupling is likely to be strongly 
channel-dependent. In particular, this dependence is to be expected if the 
coupling to excited octet baryon states occurs via intermediate excitation of 
decuplet baryons. Such excitations could be mediated by pion and kaon exchanges 
at long distances or, at shorter ranges, by the same magnetic gluon interactions 
that are responsible for the repulsive cores in the quark model \cite{oka12}. 
This mechanism for generating excited octet baryons would not operate in the 
SU(3)-singlet channel since the relevant intermediate states -- 
octet-decuplet or decuplet-decuplet -- cannot couple to an overall singlet.
It could therefore explain the absence of a repulsive core in that channel.

To summarise: a potential is not an observable. The toy model in this note
provides a warning that even quite striking features of a potential may not 
reflect the actual dynamics of the system it describes.

A final comment is that this toy model provides a very simple 
setting where Bethe-Salpeter amplitudes can be constructed 
with known phase shifts. It can therefore help to check 
the extent to which potentials extracted from these amplitudes 
do reproduce phase shifts. Indeed the model has already been used by 
Sugiura \textit{et al.}~to study the convergence of a derivative
expansion of energy-independent but nonlocal potentials \cite{sio17}. 
As noted in the Introduction, the time-dependent version of the
HAL QCD approach uses correlators that are not dominated by a single
energy eigenstate and so do not lead to Bethe-Salpeter amplitudes
with well-defined phase shifts. This is an issue for local potentials
determined by this approach, although not necessarily for nonlocal
ones. An extension of the model studies in Ref.~\cite{sio17}
to time-dependent amplitudes could shed light on how to implement
the method in a way that reproduces physical phase shifts \cite{ish17}.

\section*{Acknowledgments}
I am grateful to T. Sugiura and N. Ishii for discussions of how 
this model could be used to test their approach. 
This work was supported by the UK STFC under grants ST/J000159/1,
ST/L005794/1 and ST/P004423/1.

\end{document}